\documentclass[12pt]{article}

\def\mytitle#1{\setcounter{equation}{0}
\setcounter{footnote}{0}
\begin{flushleft}\Large\textbf{#1}\end{flushleft}
\vspace{0.27cm}}
\def\myname#1{\leftline{{\large #1}}\vspace{-0.13cm}}
\def\myplace#1#2{\small\begin{flushleft}\textit{#1}\\
\texttt{#2}\end{flushleft}}

\def\myclassification#1{\small\noindent
Pacs no : 04.20-q, 04.30-w, 04.40.Nr
       #1\vspace{0.5cm}}
\usepackage{graphicx}
\begin{document}

\mytitle{Gravitational collapse of cylindrical anisotropic fluid: A source of gravitational waves}

\vskip0.2cm \myname{Sanjukta Chakraborty\footnote{sanjuktachakraborty77 @gmail.com}}
\vskip0.2cm \myname{Subenoy Chakraborty\footnote{schakraborty@math.jdvu.ac.in}}

\myplace{Department of Mathematics, Tarakeswar Degree College, Tarakeswar, India.}{}
\myplace{Department of Mathematics, Jadavpur University, Kolkata-700 032, India.}{}

\begin{abstract}

The present work deals with dynamics of gravitational collapse with cylindrical symmetry as developed by Misner and Sharp. The interior collapsing anisotropic cylindrical perfect fluid is matched to an exterior  vacuum cylindrically symmetric space-time due to Einstein--Rosen using the Darmois matching conditions. It is found that the radial pressure of the anisotropic perfect fluid is non-zero on the boundary surface and is related to the components of shear viscosity. As a result, there is formation of gravitational waves outside the collapsing matter.\\

\end{abstract}
\myclassification{}\\
\section{Introduction}

In gravitational physics as well as in relativistic astrophysics, a well known curiosity is to know the final fate of an endless gravitational collapse. In a massive star, the inward pull of gravity is counter balanced  by the outward pressure of the nuclear fuel at the core of the star. But after completion of thermonuclear burning, the star has exhausted its nuclear fuel and as a result there is a continual gravitational collapse at the end of its life cycle.\\

Long back in 1939, Oppenheimer and snyder [1] initiated the study of gravitational collapse of a homogeneous spherical dust cloud in the frame work of general relativity. Afterwards, Joshi and Singh [2] gave an analytic study of spherically symmetric collapse of an inhomogeneous dust cloud. They concluded that end state of collapse (black hole or naked singularity) depends on initial density distribution and the radius of the star. Keeping in mind that pressure has a crucial role at the end stages of collapse, Misner and Sharp [3] studied collapse of a spherically symmetric ideal fluid with Schwarzschild space-time at the exterior of the star.\\

However, the majority of works done on this issue assume spherical symmetry.Even it is true that in the study of self-gravitating compact objects, deviations from spherical symmetry are likely to be incidental rather than basic features of the process involved. Also it is true that extensions to other kind of symmetries may provide important information about self-gravitating fluids in general.\\

In fact cylindrical systems in Einstein's theory puzzles relativists since Levi-Civita found its vacuum solution. The precise meaning of its two independent parameters is still unknown. Besides, there has been renewed interest in cylindrically symmetric sources in relation with different, classical and quantum aspects of gravitation. Such sources may serve as test bed fot numerical relativity, Quantum gravity and for probing cosmic censorship and hoop conjecture, among other important issues, and represent a natural tool to seek the physics that lies behind the two independent parameters in Levi-Civita metric.\\

Further, consideration of dissipative matter is more realistic as gravitational collapse is a highly dissipative process [4-6]. Gravitational collapse of a radiating star was considered by Chan [7] with dissipation in the form of radial heat flow and shear viscosity. He showed that anisotropy of pressure which is increased by shear viscosity, plays a significant role in the study of gravitational collapse. Subsequently, following the formulation of Misner and Sharp, Herrera and santos [5] discussed dynamical description of gravitational collapse with dissipation of energy as heat flow and radiation. Also Herrera etal [8-10]  formulated the collapse dynamics considering dissipation in the form of heat flow, radiation, shear and bulk viscosity and then coupled with causal transport equations. The same collapsing process with geometry in the form of plane symmetry has been considered recently by Sharif and Rehmat[11].\\

Herrera et al[12] for the first time claimed that if one matches a cylindrical non-dissipative fluid to an exterior containing gravitaional waves, there is a non-vanishing radial pressure on the boundary surface. Subsequently, the same authors with M.A.H. Maccallum [13]showed that the above was wrong. Also, shear free cylindrical collapse has been analyzed in great detail by Prisco et al [14] and a general study including the shearing case has been reported by Herrera et al [15] in recent past.\\

Moreover, in the context of gravitational collapse, the junction conditions due to Darmois [16] across the boundary surface of the collapsing object is essential for its study. Sharif and Ahmed [17-19] studied the junction conditions between static exterior and non-static interior with a cosmological constant and showed how collapsing process is slowed down due to the presence of the positive cosmological constant.By using Darmois junction conditions, Herrera et al [8] showed that any conformally flat cylindrically symmetric static source cannot be matched to the Levi-Civita space-time.Also, Kurita and Nakao [20] showed that naked singularity is formed along the axis of symmetry, in studying collapse of null dust in the cylindrically symmetric space-time.\\

On the other hand, to describe sources radiating gravitational waves, one must have to deviate from spherical symmetry. This fact posses an important challenge, namely: For bound sources, the exterior space time should in principle describe such a radiation. However, it is well known that no exact solution, describing gravitational radiation from bound sources, is available in closed analytic form. Only Bondi approach provides metric functions as inverse power series of the null co-ordinate and it converges very far from the source. In particular, there is no explicit exterior metric to which we could match our interior fluid distribution except in the cylindrically symmetric case, where we have the Einstein--Rosen space-time.\\

Further, in the context of gravitational waves it is generally known that a source radiating gravitational waves losses mass [21-23] as waves carry energy. But the fluid pressure of the source is not affected by radiation of gravitational waves. However, if one considers collapse of a cylindrical non-dissipative fluid with exterior containing gravitational waves, then by Darmois matching conditions there should be a non-zero pressure on the collapsing boundary surface. In the present work, we consider collapse of a cylindrical dissipative anisotropic fluid with exterior vacuum in Einstein--Rosen co-ordinates [24]. By using Darmois matching conditions, it is examined how the dissipative effect (shear viscosity) changes the anisotropic pressure and influences the collapse dynamics. The plan of the paper is as follows. The basic equations both for interior and exterior has been presented in section 2. Section 3 deals with the junction conditions on the boundary surface. Section 4 shows how the radial pressure on the boundary depends on the shear components, the metric components and radial velocity by using the junction conditions. Solution to the emitted pulse has been evaluated in section 5. Finally there is summary of the work in the section 6. \\

\section{Anisptropic fluid as interior matter distribution of collapsing cylinder}

Let us consider a cylindrical surface with its motion described by a timelike three surface $\Sigma$. So the four dimensional space-time is divided into two disjoint manifolds $M^-$ (interior) and $M^+$ (exterior) having common boundary $\Sigma$.
Using comoving coordinates, the interior manifold $M^-$ can be described by the general time dependent diagonal non rotating cylindrically symmetric metric as  
\begin{equation}
d{s_-^2}=-{A^2}d{t^2} +{B^2 }d{r^2} +{C^2}d{\phi^2} +{D^2}d{z^2}
\end{equation}\\
where A,B,C and D are functions of t and r and notationally we write $\lbrace x^{-\mu} \rbrace \equiv[t,r,\phi,z] (\mu=0,1,2,3) $.\\

For cylindrical symmetry the coordinates are restricted to the ranges:\\
$-\infty\leq t\leq +\infty,~~~r\geq 0,~~~-\infty<z<+\infty,~~~0\leq \phi \leq 2\Pi$\\

We consider the collapsing cylinder filled with anisotropic dissipative fluid (bounded by $\Sigma$) (comoving in this coordinate system) having energy-momentum tensor 

\begin{equation}
T_{\mu\nu}=(\rho+p_t){v_\mu}{v_\nu}+{p_t}g_{\mu\nu}+({p_r}-{p_t}){\chi_\mu}{\chi_\nu}-2\eta\sigma_{\mu\nu}
\end{equation}\\
where $\rho,~p_r,~p_t~ and~~\eta$ are the energy density ,the radial pressure, the tangential pressure and the coefficient of shear viscosity.
Here $v_\mu$ and $\chi_\mu$  are respectively unit time-like and space-like vectors i.e.
\begin{equation}
v_\mu v^\mu =-\chi_\mu\chi^\mu=-1~~~,~~~~\chi^\mu v_\mu=0
\end{equation}\\
and the shear tensor $\sigma_{\mu\nu}$ has the expression 
\begin{equation}
\sigma_{\mu\nu}= v_{(\mu;\nu)}+a_{(\mu}v_{\nu)}-\frac{1}{3}\Theta(g_{\mu\nu}+v_\mu v_\nu)
\end{equation}\\
where $a_\mu=v_{\mu;\nu}v^\nu $ is the acceleration vector and $\Theta=v^\mu;_\mu$ is the expansion scalar. Note that there is no explicit term of bulk viscosity in the energy-momentum tensor as its components can be absorbed in the form of radial and tangential presssures of the collapsing fluid. Let us choose in the above co-moving co-ordinates the four velocity and the unit space-like vector as 
\begin{equation}
v^\mu=A^{-1}\delta_0^\mu~~~,~~~~~~~~~~~~~~\chi^\mu=B^{-1}\delta_1^\mu
\end{equation}\\
Here the non-vanishing components of the shear tensor are  
\begin{equation}
\sigma_{11}=\frac{B^2}{3A}[\Sigma_1-\Sigma_3]~~,~~~~~~\sigma_{22}=\frac{C^2}{3A}[\Sigma_2-\Sigma_1]~~~,\sigma_{33}=\frac{D^2}{3A}[\Sigma_3-\Sigma_2]~~~and~~~~~\sigma^2=\frac{1}{6A^2}[\Sigma_1^2+\Sigma_2^2+\Sigma_3^2]
\end{equation}\\
where
\begin{equation}
 \Sigma_1=\frac{\dot{B}}{B}-\frac{\dot{C}}{C}~~,~~~\Sigma_2=\frac{\dot{C}}{C}-\frac{\dot{D}}{D}~~,~~~\Sigma_3=\frac{\dot{D}}{D}-\frac{\dot{B}}{B}
\end{equation}\\
Also the explicit form of the acceleration vector and the expansion scalar are given by 

\begin{equation}
a_1=\frac{A^\prime}{A}~~,\Theta=\frac{1}{A}(\frac{\dot{B}}{B}+\frac{\dot{C}}{C}+\frac{\dot{D}}{D})
\end{equation}\\
where, $\cdot$ $\equiv\frac{\partial}{\partial t}$~~~~~and~~~~~$^\prime$ $\equiv\frac{\partial}{\partial r}.$\\

For the metric (1) the Einstein field equations $G_{\mu\nu}=\kappa T_{\mu\nu}$ (using equations ((2)-(8))), reduces to five non-zero components, but we shall need only the following two:
$G_{11}=\kappa T_{11}$
i.e.\\
\begin{equation}
\frac{1}{A^2}[\frac{\ddot{C}}{C}+\frac{\ddot{D}}{D}+\frac{\dot{C}}{C}\frac{\dot{D}}{D}-\frac{\dot{A}}{A}\frac{\dot{C}}{C}-\frac{\dot{A}}{A}\frac{\dot{D}}{D}]-\frac{1}{B^2}[\frac{C^\prime}{C}\frac{D^\prime}{D}+\frac{A^\prime}{A}\frac{C^\prime}{C}+\frac{A^\prime}{A}\frac{D^\prime}{D}]=- \kappa [p_r-\frac{2\eta}{3A}(\Sigma_1-\Sigma_3)]
\end{equation}\\
$G_{10}=0$
\begin{equation}
-\frac{\dot{C}^\prime}{C}-\frac{\dot{D}^\prime}{D}+\frac{C^\prime}{C}\frac{\dot{B}}{B}+\frac{D^\prime}{D}\frac{\dot{B}}{B}+\frac{A^\prime}{A}\frac{\dot{C}}{C}+\frac{A^\prime}{A}\frac{\dot{D}}{D}=0
\end{equation}
The non-zero components of the energy conservation relation $T_{\nu;\mu}^\mu=0$ are given by 
\begin{equation}
\dot{\rho}+(\rho+p_r)\frac{\dot{B}}{B}+(\rho+p_t)(\frac{\dot{C}}{C}+\frac{\dot{D}}{D})=0
\end{equation}
and
\begin{equation}
p_r^\prime+(\rho+p_r)\frac{B^\prime}{B}+(p_r-p_t)(\frac{C^\prime}{C}+\frac{D^\prime}{D})=0
\end{equation}

For the exterior vacuum space-time manifold $M^+$ with $\Sigma$ as the boundary we choose the metric in Einstein -Rosen co-ordinates [20] as 
\begin{equation}
ds_+^2=-e^{2(\gamma-\psi)}(dT^2-dR^2)+e^{2\psi}dz^2+e^{-2\psi}R^2d\phi^2
\end{equation}

with $\psi=\psi(T,R)$,~~~~~~~~~~~$\gamma=\gamma(T,R)$~~~~~~~~~~and~~~$x^{+\mu}=(T,R,\phi,z)$ 

The vacuum Einstein field equations $R_{\mu\nu}^+=0$ gives the cylindrically symmetric  wave equation in an Euclidean space-time i.e. 
\begin{equation}
\psi_{,TT}-\psi_{,RR}-\frac{\psi_{,R}}{R}=0
\end{equation}
\begin{equation}
\gamma_{,T}=2R\psi_{,T}\psi_{,R}~~~~~,\gamma_{,R}=R(\psi_{,T}^2+\psi_{,R}^2)
\end{equation}

The wave equation (14) indicates possible existence of a gravitational wave field.

\section{Matching of the interior and exterior manifolds:Junction conditions}

In order to have a smooth matching of the interior and exterior manifolds on the bounding three surface $\Sigma$(not a surface layer), the necessary and sufficient conditions are that both the first and the second fundamental forms of the two manifold should be continuous across $\Sigma$ [16].
From the point of view of the interior manifold we can write down the surface $\Sigma$ as 
\begin{equation}
r-r_\Sigma=0
\end{equation}
where $r_\Sigma$ is a constant as $\Sigma$ is a comoving surface forming the boundary of the fluid. Now using (16) in (1) we have the metric on $\Sigma$ as 
\begin{equation}
ds^2\stackrel{\Sigma}{=}-d\tau^2 +C^2dz^2+D^2d\phi^2
\end{equation}
 where
\begin{equation}
 d\tau\stackrel{\Sigma}{=}Adt
\end{equation}

defines the time co-ordinate on $\Sigma$. Here, the notation $"\stackrel{\Sigma}{=}"$ implies the equality on both sides of the surface $\Sigma$. For convenience, we choose $\xi^0= \tau,\xi^2= z,~~\xi^3=\phi$ as the parameters on $\Sigma$.\\
Similarly, considering the exterior manifold the surface $\Sigma$ can be characterized as 

\begin{equation}
R-R_\Sigma(T)=0
\end{equation}
  Using (19) in (13), the metric on $\Sigma$ can again be written as 
\begin{equation}
ds^2\stackrel{\Sigma}{=}-e^{2(\gamma-\psi)}	[1-(\frac{dR}{dT})^2]dT^2+e^{2\psi}dz^2+e^{-2\psi}R^2d\phi^2
\end{equation}

So for smooth matching of the 1st fundamental form, comparing (17) and (20)we have

\begin{equation}
d\tau\stackrel{\Sigma}{=}e^{\gamma-\psi}[1-(\frac{dR}{dT})^2]^{1/2}dT\stackrel{\Sigma}{=}Adt
\end{equation}

\begin{equation}
C\stackrel{\Sigma}{=}e^\psi
\end{equation}

\begin{equation}
D\stackrel{\Sigma}{=}e^{-\psi}R
\end{equation}

Note that as T is time-like coordinate so $[1-(\frac{dR}{dT})^2]>0$ on $\Sigma$\\ 

The continuity of the second fundamental form on $\Sigma$ i.e., $K_{ij}d\xi^id\xi^j$ implies the continuity of the extrinsic curvature over the hypersurface [16]
 
\begin{equation}
[K_{ij}]\equiv K_{ij}^+ -K_{ij}^-=0
\end{equation}

where $K_{ij}^\pm$ is given by

\begin{equation}
K_{ij}^\pm=-n_\sigma^\pm[\frac{\partial^2x_\pm ^\sigma}{\partial\xi^i \partial\xi^j}+\Gamma_{\mu\nu}^\sigma \frac{\partial x_\pm^\mu}{\partial\xi^i}\frac{\partial x_\pm^\nu}{\partial\xi^j}],~~~(\sigma,~\mu,~\nu~=0,1,2,3)
\end{equation}

Here $n_\sigma^\pm$ are the components of the outward unit normal to the hypersurface in the co-ordinates $x^{\pm\mu}$ and the christoffel symbols are to be calculated from the appropriate exterior or interior metric .\\

The explicit components of the unit normal vectors are of the form

\begin{equation}
n_\sigma^- \stackrel{\Sigma}{=}(0,B_\Sigma,0,0)~~\\
and~~~~\\
n_\sigma^+ \stackrel{\Sigma}{=}e^{2(\gamma-\psi)}(-R_\tau,T_\tau,0,0)
\end{equation}\\
where suffix $\tau$ indicates differentiation w.r.t. $\tau$.\\

Thus from the continuity of $K_{00}^-\stackrel{\Sigma}{=}K_{00}^+$ we have 

\begin{equation}
e^{2(\gamma-\psi)}[T_{\tau\tau}R_\tau-R_{\tau\tau}T_\tau-(T_\tau^2-R_\tau^2)[R_\tau(\gamma_{,T}-\psi_{,T})+T_\tau(\gamma_{,R}-\psi_{,R})]]\stackrel{\Sigma}{=}-\frac{A^\prime}{AB}
\end{equation}\\
Similarly, from the equality of $K_{22}^-\stackrel{\Sigma}{=}K_{22}^+$ and $K_{33}^-\stackrel{\Sigma}{=}K_{33}^+$  we have 

\begin{equation}
e^{2\psi}(R_\tau \psi_{,T}+T_\tau \psi_{,R})\stackrel{\Sigma}{=}\frac{C{C}^\prime}{B}
\end{equation}\\
and\\
\begin{equation}
e^{-2\psi}R^2 (R_\tau \psi_{,T}+T_\tau \psi_{,R}-\frac{T_\tau}{R})\stackrel{\Sigma}{=}\frac{-D{D}^\prime}{B}
\end{equation}\\

\section{Deductions from the matching conditions}
In this section, we shall write down the matching conditions (formulated in the previous section) in a concise form by using the interior and exterior field equations.
From equation (21) we write 
\begin{equation}
T_\tau^2 -R_\tau^2\stackrel{\Sigma}{=}e^{2(\psi-\gamma)}		
\end{equation}\\
Combining the matching conditions (22) and (23) and using (18) we have 
\begin{equation}
R_\tau\stackrel{\Sigma}{=}\frac{(CD),t}{A}
\end{equation}\\
Also from the continuity of the extrinsic curvature $(K_{22}^-=K_{22}^+, K_{33}^-=K_{33}^+)$ i.e. equations (28) and (29) and using equations (22) and (23) we obtain 
\begin{equation}
T_\tau=\frac{1}{B}(CD),r
\end{equation}\\
Now differentiating (31) and (32) with respect to $\tau$ and using the Einstein field equations (9) and (10) we get 

\begin{eqnarray}
 T_{\tau\tau}R_\tau-T_\tau R_{\tau\tau}&=&\frac{1}{A^3 B}{A}^\prime(\dot{C}D+C\dot{D})^2+\frac{1}{A^2 B}(D{D}^\prime\dot{C}^2+C{C}^\prime\dot{D}^2)+ \frac{CD}{B}\kappa({C}^\prime D+C{D}^\prime)[p_r-\frac{2\eta}{3A}(\Sigma_1 -\Sigma_3)]
\nonumber
\\
&-&\frac{1}{B^3}({C}^\prime D+C{D}^\prime)({C}^\prime{D}^\prime+\frac{{A}^\prime{C}^\prime D}{A}+\frac{{A}^\prime{D}^\prime C}{A})
\end{eqnarray}

Using the matching condition (22) in (30) and using (31) and (32) we have on simplification 

\begin{equation}
\psi_{,T}(T_{\tau} ^2 -R_{\tau} ^2)=\frac{1}{AB}(\dot{C} D^\prime- C^ \prime \dot{D})
\end{equation}

Substituting (33)and (34) in the continuity of $K_{00}$ (.i.e. eq (27)) and using the field equations (15) we obtain 

\begin{equation}
(p_r)_{eff}= p_r -\frac{2\eta}{3A}(\Sigma_1-\Sigma_3) \stackrel{\Sigma} {=} 0
\end{equation}

The above result shows that the radial pressure $p_r$ on the boundary surface $\Sigma$ of the collapsing perfect fluid is non-zero due to the presence of the shear viscosity. Note that in absence of dissipation, our result agrees with that in reference [13]. In the present scenario, the effective radial pressure vanishes on the boundary. This is expected from equivalence principle which forbids any local characterization of gravitational energy. In fact, the pressure is a local scalar, and therefore, a non-vanishing pressure associated with gravitational radiation would imply a local measurement of gravitational interaction, which in turn would contradict the equivalence principle.

\section{Solution to the emitted pulse} 
Suppose a collapsing cylinder remains static for a small time and then it gradually contracts. As a result, a sharp pulse of radiation will emit outward from the axis and we have [25] 
\begin{equation}
\psi=\frac{1}{2\Pi}\int^{T-R}_{-\infty}\frac{f(T^\prime)dT^\prime}{[(T-T^\prime)^2-R^2]^\frac{1}{2}}+\psi_{st}
\end{equation}

where $\psi_{st}$ is the usual Levi -Civita static solution [26], $f(T)$ indicates the strength of the wave source and is chosen as dirac delta function i.e.
\begin{equation}
f(T)=f_0\delta(T),
\end{equation}
 $f_0$ is a constant. As a result, we have 

\begin{equation}
\psi=\left\{
\begin{array}{ll}
\psi_{st} & \mbox{ for } T{< }R\\
\frac{f_0}{2\Pi(T^2-R^2)^{\frac{1}{2}}}+\psi_{st} & \mbox{ for } T{>}R
\end{array}
\right\}
\end{equation}  

Note that $\psi$ and its derivative is smooth everywhere except at the wave front i.e. the surface $T=R$. Further, for sufficient large $T, \psi_T, \psi_R$ (which is related to the rate of change of the $C$ energy) is positive on the boundary surface $\Sigma$. However, in the literature [23] it has been shown to be negative only for large $R$. \\

Finally, if we consider the exterior manifold to be static (i.e. $\psi=\psi(R)$) then from equations (22), (23), (31) and (32) and the continuity of $K_{22}$ and $K_{33}$ we see that $C\stackrel{\Sigma}{=}C(r)$ and $D\stackrel{\Sigma}{=}D(r)$ i.e. $R\stackrel{\Sigma}{=}$ constant which shows that there will no longer be any collapse. This result is consistent with the known result that the static exterior space-time (i.e.Levi-civita) can not match to a collapsing cylinder with matter source. Also this result has been shown in [27] for collapsing dust cylinder. \\

\section{Summary of the work:}
The present work  studies the collapse dynamics of an anisotropic (dissipative) cylindrical perfect fluid along the line of formulation of Misner and Sharp. Here the dissipation is due to shear viscosity. The exterior manifold is chosen as vacuum cylindrically symmetric space-time due to Einstein and Rosen. The matching of the interior and exterior manifolds over the bounding surface is done following Darmois matching conditions. From the junction conditions it has been shown that the surface pressure is non-zero only due to the dissipative components. However, the effective radial pressure on the boundary vanishes as expected from equivalence principle and it agrees with the result of Herrera and collaborators [13]. Also, it should be mentioned that the result of Bondi et al [28] that there will no longer be any radiation of gravitational waves if we have collapsing dust cylinder, is supported by the present work. Finally, one may conclude that a collapsing cylindrical matter dissipative in nature always produce gravitational waves outside the matter distribution and as a result there will be non-zoro radial pressure on the bounding surface. \\

\section{Acknowledgement}
One of the author (SC) is thankful to UGC-DRS programme, Department of Mathematics, Jadavpur University. The authors are thankful to the reviewer for pointing out the error and for valuable suggestions. The authors are also grateful to L. Herrera and N.O. Santos for their help in detecting the error.

\end{document}